\documentclass[12pt,oneside]{article}
\usepackage{epsfig, cite}
\usepackage{color}
\usepackage{ifthen}
\usepackage{graphicx}

\begin{document}
\thispagestyle{empty}
\setcounter{page}{0}
\renewcommand{\theequation}{\thesection.\arabic{equation}}

\vspace{2cm}

\begin{center}
{\bf THE COSMOLOGICAL CONSTANT\\ AND THE STRING LANDSCAPE}

\vspace{1.4cm}

JOSEPH POLCHINSKI

\vspace{.2cm}

{\em Kavli Institute for Theoretical Physics} \\
{\em University of California} \\
{\em Santa Barbara, CA 93106-4030} \\
\end{center}

\vspace{-.1cm}

\centerline{{\tt joep@kitp.ucsb.edu}}
\vspace{1cm}
\centerline{ABSTRACT}

\vspace{- 4 mm}  

\begin{quote}\small
Theories of the cosmological constant fall into two classes, those in which the vacuum energy is fixed by the fundamental theory and those in which it is adjustable in some way.  For each class we discuss key challenges.  The string theory landscape is an example of an adjustment mechanism.  We discuss the status of this idea, and future directions.

Rapporteur talk at the 23rd Solvay Conference
in Physics, December, 2005.
\end{quote}
\baselineskip18pt
\noindent

\vspace{5mm}

\newpage

\setcounter{equation}{0}
\section{The cosmological constant}
I would like to start by drawing a parallel to an earlier meeting --- not a Solvay conference, but the 1947 Shelter Island conference.  In both cases a constant of nature was at the center of discussions.  In each case theory gave an unreasonably large or infinite value for the constant, which had therefore been assumed to vanish for reasons not yet understood, but in each case experiment or observation had recently found a nonzero value.  At Shelter Island that constant was the Lamb shift, and here it is the cosmological constant.  But there the parallel ends: at Shelter Island, the famous reaction was ``the Lamb shift is nonzero, therefore we can calculate it,'' while today we hear ``the cosmological constant is nonzero, therefore we can calculate {\em nothing.}''  Of course this is an overstatement, but it is clear that the observation of an apparent cosmological constant has catalyzed a crisis, a new discussion of the extent to which fundamental physics is predictable.  This is the main subject of this report.

In the first half of my talk I will review why the cosmological constant problem is so hard.  Of course this is something that we have all thought about, and there are major reviews.\footnote{For a classic review see~\cite{Weinberg:1988cp}.  For more recent reviews that include the observational situation and some theoretical ideas see \cite{Carroll:2000fy, Padmanabhan:2002ji}.  A recent review of theoretical ideas is~\cite{Nobbenhuis:2004wn}.  My report is not intended as a comprehensive review of either the cosmological problem or of the landscape, either of which would be a large undertaking, but a discussion of a few key issues in each case.}  However, given the central importance of the question, and the flow of new ideas largely stimulated by the observation of a nonzero value, we should revisit this.  One of my main points is that, while the number of proposed solutions is large, there is a rather small number of principles and litmus tests that rule out the great majority of them.

In recent years the cosmological constant has become three problems:
\begin{enumerate}
\item
Why the cosmological constant is not large.
\item
Why it is not zero.
\item
Why it is comparable to the matter energy density {\it now} (cosmic coincidence).
\end{enumerate}
I will focus primarily on the first question --- this is hard enough! --- and so the question of whether the dark energy might be something other than a cosmological constant will not be central.

In trying to understand why the vacuum does not gravitate, it is useful to distinguish two kinds of theory:
\begin{enumerate}
\item
Those in which the energy density of the vacuum is more-or-less uniquely determined by the underlying theory.
\item
Those in which it is not uniquely determined but is adjustable in some way.
\end{enumerate}
I will discuss these in turn.

\subsection{Fixed-$\Lambda$ theories}

The basic problem here is that we know that our vacuum is a rather nontrivial state, and we can identify several contributions to its energy density that are of the order of particle physics scales.  It is sufficient to focus on one of them; let us choose the electron zero point energy, since we know a lot about electrons.  In particular, they are weakly coupled and pointlike up to an energy scale $M$ of at least 100 GeV.  Thus we can calculate the electron zero point energy up to this scale from the graphs of Fig.~1~\cite{Coleman:1973jx},
\begin{equation}
\rho_V = O(M^4) + O(M^2 m_e^2) + O(m_e^4 \ln M/m_e)\ , \label{cwzpe}
\end{equation}
which is at least 55 orders of magnitude too large.
\begin{figure}[h]
\begin{center}
\leavevmode
\epsfbox{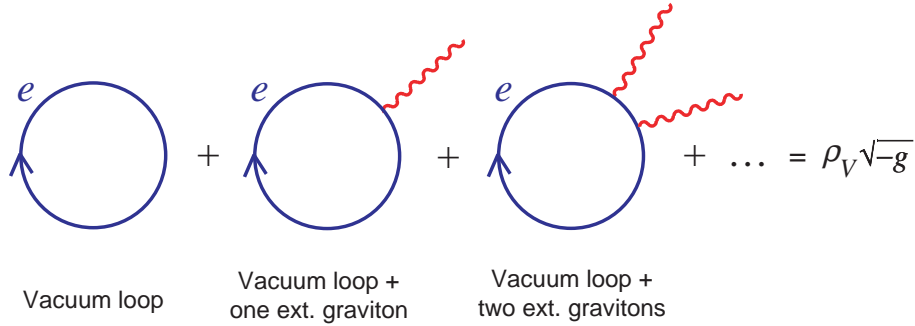}
\end{center}
\caption{An electron vacuum loop and its coupling to external gravitons generate an effective cosmological constant.}
\end{figure}

So we must understand why this contribution actually vanishes, or is cancelled.  To sharpen the issue, we know that electron vacuum energy does gravitate in some situations.
\begin{figure}[h]
\begin{center}
\leavevmode
\epsfbox{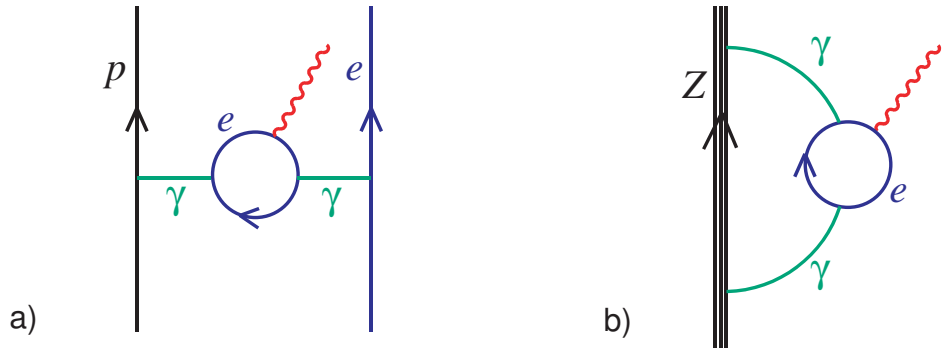}
\end{center}
\caption{a) Vacuum polarization contribution to the Lamb shift, coupled to an external graviton.  b) A loop correction to the electrostatic energy of a nucleus, coupled to an external graviton.}
\end{figure}
Fig.~2a shows the vacuum polarization contribution to the famous Lamb shift, now coupled to an external graviton.  Since this is known to give a nonzero contribution to the energy of the atom, the equivalence principle requires that it couple to gravity.  The Lamb shift is very small so one might entertain the possibility of a violation of the equivalence principle, but this is a red herring, as there are many larger effects of the same type.  

One of these is shown in Fig.~2b, a loop correction to the electrostatic energy of the nucleus.  Aluminum and platinum have the same ratio of gravitational to inertial mass to one part in $10^{12}$~\cite{eotvos1,eotvos2}.  The nuclear electrostatic energy is roughly $10^{-3}$ of the rest energy in aluminum and $3 \times 10^{-3}$ in platinum.  Thus we can say that this energy satisfies the equivalence principle to one part in $10^{9}$.  The loop graph shifts the electrostatic energy by an amount of relative order ${\alpha}\ln(m_e R_{\rm nuc})/{4\pi} \sim 10^{-3}$ due to the running of the electromagnetic coupling.  Thus we know to a precision of one part in $10^6$ that the effect shown in Fig.~2b actually exists.  In fact, the effect becomes much larger if we consider quark loops rather than electrons, and we do not need precision experiments to show that virtual quarks gravitate, but we stick with electrons because they are cleaner~\cite{Bagger:2000iu}.

We can think of Fig.~2 to good approximation as representing the shift of the electron zero point energy in the environment of the atom or the nucleus.  Thus we must understand why the zero point energy gravitates in these environments and not in vacuum, again given that our vacuum is a rather complicated state in terms of the underlying fields.  Further, if one thinks one has an answer to this, there is another challenge: why does this cancellation occur in our particular vacuum state, and not, say, in the more symmetric $SU(2) \times U(1)$ invariant state of the weak interaction?  It cannot vanish in both because the electron mass is zero in the symmetric state and not in ours, and the subleading terms in the vacuum energy~(\ref{cwzpe}) --- which are still much larger than the observed $\rho_V$ --- depend on this mass.  Indeed, this dependence is a major contribution to the Higgs potential (though it is the top quark loop rather than the electron that dominates), and they play an important role in Higgs phenomenology.

I am not going to prove that there is no mechanism that can pass these tests.  Indeed, it would be counterproductive to do so, because the most precise no-go theorems often have the most interesting and unexpected failure modes.  Rather, I am going to illustrate their application to one interesting class of ideas.

Attempts to resolve the Higgs naturalness problem have centered on two mechanisms, supersymmetry and compositeness (technicolor).  In the case of the cosmological constant much attention has been given to the effects of supersymmetry, but what about compositeness, technigravity?  If the graviton were composite at a scale right around the limit of Cavendish experiments, roughly 100 microns, would this not cut off the zero point energy and leave a remainder of order $(100 \,\mu)^{-4}$, just the observed value~\cite{Beane:1997it,Sundrum:1997js}?  Further this makes a strong prediction, that deviations from the inverse square law will soon be seen.  

In fact, it can't be that simple.  When we measure the gravitational force in Cavendish experiments, the graviton wavelength is around $100 \,\mu$.  When we measure the cosmological constant, the graviton wavelength is around the Hubble scale, so there is no direct connection between the two.  Moreover, we already know, from the discussion of Fig.~2, that the coupling of gravity to off-shell electrons is unsuppressed over a range of scales in between $100 \,\mu$ and the Hubble scale, so whatever is affecting the short-distance behavior of gravity is not affecting longer scales.  We can also think about this as follows: even if the graviton were composite one would not expect the graphs of Fig.~1 to be affected, because all external fields are much softer than $100 \,\mu$.  In order to be sensitive to the internal structure of a particle we need a hard scattering process, in which there is a large momentum transfer to the particle~\cite{Rutherford}.
Further, the large compact dimension models provide an example where gravity is modified at short distance, but the electron zero point loop is not cut off at that scale.  Thus there is no reason, aside from numerology, to expect a connection between the observed vacuum energy and modifications of the gravitational force law.

Ref.~\cite{Sundrum:2003tb} tries to push the idea further, defining an effective theory of `fat gravity' that would pass the necessary tests.  This is a worthwhile exercise, but it shows just how hard it is.  In order that the vacuum does not gravitate but the Lamb shift and nuclear loops do, fat gravity imposes special rule for vacuum graphs.  The matter path integral, at fixed metric, is doubly nonlocal: there is a UV cutoff around $100 \,\mu$, and in order to know how to treat a given momentum integral we have to look at the topology of the whole graph in which it is contained.  Since the cosmological constant problem really arises only because we know that some aspects of physics are indeed local to a much shorter scale, it is necessary to derive the rules of fat gravity from a more local starting point, which seems like a tall order.  To put this another way, let us apply our first litmus test: what in fat gravity distinguishes the environment of the nucleus from the environment of our vacuum?  The distinction is by fiat.  But locality tells us that that the laws of physics are simple when written in terms of local Standard Model fields.  Our vacuum has a very complicated expression in terms of such fields, so the rules of fat gravity do not satisfy the local simplicity principle.

The nonlocality becomes sharper when we look at the second question, that is, for which vacuum is the cosmological constant small?  The rule  given is that it is the one of lowest energy.  This sounds  simple enough, but consider a potential with two widely separated local minima.  In order to know how strongly to couple to vacuum A, the graviton must also calculate the energy of vacuum B (and of every other point in field space), and if it is smaller take the difference.  Field theory, even in some quasilocal form, can't do this --- there are not enough degrees of freedom to do the calculation.  If the system is in state A, the dynamics at some distant point in field space is irrelevant.  Effectively we would need a computer sitting at every spacetime point, simulating all possible vacua of the theory.  Later we will mention a context in which this actually happens, but it is explicitly nonlocal in a strong way.

The failure of short-distant modifications of gravity suggests another strategy: modify gravity at very long distances, comparable to the current Hubble scale, so that it does not couple to vacuum energy at that scale.  There is of course a large literature on long-distance modifications of gravity; here I will just point out one problem.  If we have zero point energy up to a cutoff of $M \sim 100$ GeV, the radius of curvature of spacetime will be of order $M_{\rm P}/M^2$, roughly a meter.  So modifications of gravity at much longer distances do not solve the problem, the universe curls up long before it knows about the modification.  It is possible that the spacetime curvature decays away on a timescale set by the long-distance modifications, but this would imply a large and uncanceled cosmological constant until quite recently.\footnote{One might consider models where this decay occurs in an epoch before the normal Big Bang, but this runs into the empty universe problem to be discussed in Sec.~1.2.}  These problems have already been discussed in Ref.~\cite{Arkani-Hamed:2002fu}, which argues that long distance modifications of gravity can account for the cosmological constant only in combination with acausality.

In another direction, it is tempting to look for some sort of feedback mechanism, where the energies from different scales add up in a way that causes the sum to evolve toward zero.  The problem is that only gravity can measure the cosmological constant --- this term in the action depends only on the metric --- so that the contribution from a scale $M$ is only observed at a much lower scale $M^2/M_{\rm P}$, and we cannot cancel $O(M^4)$ against $O(M^8/M_{\rm P}^4)$.  In another language, the cosmological constant has scaling dimension zero and we want to increase it to dimension greater than four; but gravity is clearly classical over a wide range of scales so there is no possibility of this.

Again, there is no proof that some  fixed-$\Lambda$ solution does not exist; perhaps our discussion will spur some reader into looking at the problem in a new way.
In fact there is at least one idea that is consistent with our tests: a symmetry {\it energy} $\to -${\it energy}.  This requires a doubling of degrees of freedom, so the electron loop is cancelled by a mirror loop of negative energy.  This idea is discussed as an exact symmetry in ref.~\cite{Linde:1988ws} and as an approximate symmetry not applying to gravity in ref.~\cite{Linde:1984ir,Kaplan:2005rr}; the two cases are rather different because the coordinate invariance is doubled in the first.  It might be that either can be made to work at a technical level, and the reader is invited to explore them further, but I will take this as a cue to move on to the next set of ideas.

\subsection{Adjustable-$\Lambda$ theories}

Many different mechanisms have been put forward that would avert the problems of the previous section by allowing the cosmological constant to adjust in some way; that is, the vacuum energy seen in the low energy theory is not uniquely determined by the underlying dynamics.  A partial list of ideas includes
\begin{itemize}
\item
Unimodular gravity~(see Ref.~\cite{Weinberg:1988cp} for a discussion of the history of this idea, which in one form goes back to Einstein).
\item
Nonpropagating four-form field strengths~\cite{Aurilia:1980xj,Duff:1980qv}.
\item
Scalar potentials with many minima~\cite{Bardeen:1982wj,Abbott:1984qf,Banks:1991mb}.
\item
A rolling scalar with a nearly flat potential~\cite{Banks:1984cw,Linde:1986dq}; the potential must be very flat in order that the vacuum energy be constant on shorter than cosmological times, and it must have a very long range to span the necessary range of energies.
\item
Spacetime wormholes~\cite{Coleman:1988cy,Giddings:1988cx,Banks:1988je,Coleman:1988tj}.
\item
The metastable vacua of string theory~\cite{Sakharov:1984ir,Linde:2005ht,Smolin:1994vb,Bousso:2000xa,Feng:2000if,Silverstein:2001xn,Kachru:2003aw,Susskind:2003kw}.
\item
Self-tuning (an undetermined boundary condition at a singularity in the compact dimensions)~\cite{Arkani-Hamed:2000eg,Kachru:2000hf}.
\item
Explicit tuning (i.e. an underlying theory with at least one free parameter not determined by any principle).
\end{itemize}
The possible values of $\rho_V$ must either be continuous, or form a sufficiently dense discretuum that at least one value is as small as observed.  It is important to note that {\it zero} cannot be a minimum, or otherwise special, in the range of allowed values.  The point is that the electron zero point energy, among other things, gives an additive shift to the vacuum energy; if the minimum value for $\rho_V$ were zero we would have to revert to the previous section and ask what it is cancels the energy in this true vacuum.

In this adjustable scenario, the question is, what is the mechanism by which the actual small value seen in nature is selected?  In fact, one can identify a number of superficially promising ideas:
\begin{itemize}
\item
The Hartle-Hawking wavefunction~\cite{Hartle:1983ai}
\begin{equation}
|\Psi_{\rm HH}|^2 = e^{3/8 G^2 \rho_V}
\end{equation}
strongly favors the smallest positive value of the cosmological constant~\cite{Baum:1984mc,Hawking:1984hk}.
\item
The de Sitter entropy~\cite{Gibbons:1977mu}
\begin{equation}
e^S = e^{3/8 G^2 \rho_V} = |\Psi_{\rm HH}|^2 
\end{equation}
would have the same effect, and suggests that the Hartle-Hawking wavefunction has some statistical interpretation in terms of the sytem exploring all possible states.
\item
The Coleman-de Luccia amplitude~\cite{Coleman:1980aw} for tunneling from positive to negative cosmological constant vanishes for some parameter range, so the universe would be stuck in the state of smallest positive energy density~\cite{Abbott:1984qf,Brown:1988kg}.
\end{itemize}

These ideas are all tantalizing --- they are tantalizing in the same way that supersymmetry is tantalizing as a solution to the cosmological constant problem.  That is, they are elegant explanations for why the cosmological constant might be small or zero under some conditions, but not in our particular rather messy universe.  Supersymmetry would explain a vanishing cosmological constant in a sufficiently supersymmetric universe, and these mechanisms would explain why it vanishes in an {\it empty} universe.

To see the problem, note first that the above mechanisms all involve gravitational dynamics in some way, the response of the metric to the vacuum energy.  This is as it must be, because again only gravity can measure the cosmological constant.  The problem is that in our universe the cosmological constant became dynamically important only recently.  At a redshift of a few the cosmological constant was much smaller than the matter density, and so unmeasurable by gravity; at the time of nucleosynthesis (which is probably the latest that a tunneling could have taken place) today's cosmological constant would have been totally swamped by the matter and radiation densities, and there is no way that these gravitational mechanisms could have selected for it.\footnote{This might appear to leave open the possibility that the vacuum energy is at all times of the same order as the matter/radiation density.  Leaving aside the question of how this would appear phenomenologically as a cosmological constant, the simplest way to see that this does not really address the problem is to note that as the matter energy goes to zero at late times then so will the vacuum energy: this violates the principle that zero is not a special value.  By contrast, the dynamical mechanisms above all operate for a $\rho_V$-spectrum that extends to negative values.}  This is the basic problem with dynamical selection mechanisms: only gravity can measure $\rho_V$, and it became possible for it to do so only in very recent cosmological times.  These mechanisms can act on the cosmological constant only if matter is essentially absent.

Another selection principle sometimes put forward is `existence of a static solution;'  this comes up especially in the context of the self-tuning solutions.  As a toy illustration, one might imagine that some symmetry acting on a scalar $\phi$ forced $\rho_V$ to appear only in the form
$\rho_V e^{\phi}$.\footnote{Aside from the issue discussed in the text, it is difficult to find true examples of such scaling.  For example, such a form arises at string tree level~\cite{Witten:1985xb}, but it is not protected against loop corrections.  An exact but spontaneously broken scale invariance might appear to give this form (E. Rabinovici, private communication), but in that case a Weyl transformation removes $\phi$ from both the gravitational action and the potential.  The scaling in the self-tuning singularity models receives various corrections; the six-dimensional models are argued to be better than the five dimensional ones in this regard (C. Burgess and F. Quevedo, private communication).}
If we require the existence of a static solution for $\phi$ then we must have $\rho_V = 0$. Of course this seems like cheating; indeed, if we can require a static solution then why not just require a flat solution, and get $\rho_V = 0$ in one step?  In fact these are cheating because they suffer from the same kind of flaw as the dynamical ideas.  In order to know that our solution is static on a scale of say $10^{10}$ years, we must watch the universe for this period of time!  The dynamics in the very early universe, at which time the selection was presumably made, have no way to select for such a solution: the early universe was in a highly nonstatic state full of matter and energy.

Of course these arguments are not conclusive, and indeed Steinhardt's talk presents a nonstandard cyclic cosmological history that evades the above no-go argument (see also Ref.~\cite{Rubakov:1999aq}).  If one accepts its various dynamical assumptions, this may be a technically natural solution to the cosmological constant problem.  Essentially one needs a mechanism to fill the empty vacuum with energy after its cosmological constant has relaxed to near zero; it is not clear that this is in fact possible.

In the course of trying to find selection mechanisms, one is struck by the fact that, while it is difficult to {select} for a single vacuum of small cosmological constant, it is extremely easy to identify mechanisms that will populate {\it all} possible vacua --- either sequentially in time, as branches of the wavefunction of the universe, or as different patches in an enormous spatial volume.  Indeed, this last mechanism is difficult to evade, if the many vacua are metastable: inflation and tunneling, two robust physical processes, will inevitably populate them all~\cite{Linde:1983gd,Vilenkin:1983xq,Linde:1986fd}\cite{Bousso:2000xa}.

But this is all that is needed!  Any observer in such a theory will see a cosmological constant that is unnaturally small; that is, it must be much smaller than the matter and energy densities over an extended period of the history of the universe.  The existence of any complex structures requires that there be many `cycles' and many `bits': the lifetime of the universe must be large in units of the fundamental time scale, and there must be many degrees of freedom in interaction.  A large negative cosmological constant forces the universe to collapse to too soon; a large positive cosmological constant causes all matter to disperse.  This is of course the argument made precise by Weinberg~\cite{Weinberg:1987dv}, here in a rather minimal and prior-free form.\footnote{Models incorporating anthropic selection (and the basic problem with dynamical selection) were earlier discussed in Refs.~\cite{Linde:1984ir,Banks:1984cw,Linde:1986dq}.
For further reviews see Refs.~\cite{Weinberg:1988cp,Vilenkin:2004fj,Weinberg:2005fh}.}

Thus we meet the anthropic principle.   Of course, the anthropic principle is in some sense a tautology: we must live where we can live.\footnote{Natural selection is a tautology in much the same sense: survivors survive.  But in combination with a mechanism of populating a spectrum of universes or genotypes, these `tautologies' acquire great power.}  There is no avoiding the fact that anthropic selection must operate.  The real question is, {is there any} scientific {reason to expect that some additional selection mechanism is operating?}

Staying for now with the cosmological constant (other parameters will be discussed later), the obvious puzzle is the fact that the cosmological constant is an order of magnitude smaller than the most likely anthropic value.  This is an important issue, but to overly dwell on it reminds me of Galileo's reaction to criticism of his ideas because a heavier ball landed slightly before a lighter one (whereas Aristotle's theory predicted a much larger discrepancy):
\begin{quotation}
Behind those two inches you want to hide Aristotle's ninety-nine {\it braccia} [arm lengths] and, speaking only of my tiny error, remain silent about his enormous mistake.
\end{quotation}
 The order of magnitude here is the two inches of wind resistance, the ninety-nine {\it braccia} are the 60 or 120 orders of magnitude by which most or all other proposals miss.  This order of magnitude may simply be a 1.5-sigma fluctuation, or it may reflect our current ignorance of the measure on the space of vacua.

If there is a selection mechanism, it must be rather special.  It must evade the general difficulties outlined above, and it must select a value that is {\it almost exactly the same} as that selected by the anthropic principle, differing by one order of magnitude out of 120.  Occam's razor would suggest that two such mechanisms be replaced by one --- the unavoidable, tautological, one.  Thus, we should seriously consider the possibility that there is no other selection mechanism significantly constraining the cosmological constant.  Equally, we should not stop searching for such a further principle, but I think one must admit that the strongest reason for expecting to find it is not a scientific argument but a psychological one:\footnote{Again, the Darwinian analogy is notable.} we wish fundamental theory to be as predictive as we have long assumed it would be.  

The anthropic argument is not without predictive power.  We can identify a list of post- or pre-dictions, circa 1987:
\begin{enumerate}
\item
The cosmological constant is not large.
\item
The cosmological constant is not zero.
\item
The cosmological constant is similar in order of magnitude to the matter density.
\item
As the theory of quantum gravity is better understood, it will provide a microphysics in which the cosmological constant is not fixed but environmental; if this takes discrete values these must be extremely dense in Planck units.
\item
Other constants of nature may show evidence of anthropic constraints.
\end{enumerate}
Items 2 and 3 are the second and third parts of the cosmological constant problem; we did not set out to solve them, but in fact they were solved before they were known to be problems --- they are predictions.  Item~4 will be discussed in the second half of the talk, in the context of string theory.  Item~5 is difficult to evaluate, but serious arguments to this effect have long been made, and they should not be dismissed out of hand.

Let us close this half of the talk with one other perspective.  The cosmological constant problem appears to require some form of UV/IR feedback, because the cosmological constant can only be measured at long distances or late times, yet this must act back on the Lagrangian determined at short distance or early times.  We can list a few candidates for such a mechanism:
\begin{itemize}
\item
String theory contains many examples of UV/IR mixing, such as the world-sheet duality relating IR poles in one channel of an amplitude to the sum over massive states in another channel, and the radius-energy relation of AdS/CFT duality.  Thus far however, this is yet one more tantalizing idea but with no known implications for the vacuum energy.
\item 
Bilocal interactions.  The exact {\it energy} $ \to -${\it energy} symmetry~\cite{Linde:1988ws} and the wormhole solution~\cite{Banks:1988je,Coleman:1988tj} put every point of our universe in contact with every point of another.  This ties in with our earlier remarks about the computational power of quantum field theory: here the calculation of the true vacuum energy is done in the entire volume of the second spacetime.
\item
The anthropic principle.  Life, an IR phenomenon, constrains the coupling constants, which are UV quantities.
\item
A final state condition.  At several points --- in the long distance modification of gravity, and in the dynamical mechanisms --- things would have gone better if we supposed that there were boundary conditions imposed in the future and not just initially.  Later we will encounter one context in which this might occur.
\end{itemize}

To conclude, we have identified one robust framework for understanding the vacuum energy:
(1) Stuff gravitates, and the vacuum is full of stuff.  (2) Therefore the vacuum energy must have some way to adjust.  (3) It is difficult for the adjustment to select a definite small value for the vacuum energy, but it is easy to access all values, and this, within an order of magnitude, accounts for what we see in nature.  We have also identified a number of other possible hints and openings, which may lead the reader in other directions.

\setcounter{equation}{0}
\section{The string landscape}

\subsection{Constructions}

Now let us ask where string theory fits into the previous discussion.  In ten dimensions the theory has no free parameters, but once we compactify, each nonsupersymmetric vacuum will have a different $\rho_V$.  It seems clear that the cosmological constant cannot vary continuously.  Proposed mechanisms for such variation have included nondynamical form fields and a boundary condition at a singularity, but the former are constrained by a Dirac quantization condition, and the latter will undoubtedly become discrete once the internal dynamics of the `singularity' are taken into account.  (A rolling scalar with a rather flat potential might provide some effective continuous variation, but the range of such a scalar is very limited in string theory).

Given a discrete spectrum, is there a dense enough set of states to account for the cosmological constant that we see, at least $10^{60}$ with TeV scale supersymmetry breaking or $10^{120}$ with Planck scale breaking?\footnote{These numbers must be larger if the probability distribution has significant fluctuations as recently argued in Ref.~\cite{Schwartz-Perlov:2006hi}.}  The current understanding, in particular the work of KKLT~\cite{Kachru:2003aw}, suggests the existence of a large number of metastable states giving rise to a dense discretuum near $\rho_V = 0$.  A very large degree of metastability is not surprising in complicated dynamical systems --- consider the enormous number of metastable compounds found in nature.  As a related example, given 500 protons, 500 neutrons, and 500 electrons, how many very long-lived bound states are there?  A rough estimate would be the number of partitions of 500, separating the protons into groups and then assigning the same number of neutrons and electrons to each group; there is some overcounting and some undercounting here, but the estimate should be roughly correct,
\begin{equation}
P(n) \sim \frac{1}{4n\sqrt{3}} e^{\pi \sqrt{2n/3}}\ , \quad P(500) \sim 10^{22}\ .
\end{equation}
The number of metastable states grows rapidly with the number of degrees of freedom.

In string theory, replace {\it protons}, {\it neutrons}, and {\it electrons} with {\it handles}, {\it fluxes}, and {\it branes}.  There are processes by which each of these elements can form or decay, so it seems likely that most or all of the nonsupersymmetric vacua are unstable, and the space of vacua is largely or completely connected.  Thus all states will be populated by eternal inflation, if any of the de Sitter states is.  The states of positive $\rho_V$ would also be populated by any sort of tunneling from nothing (if this is really a distinct process), since one can take the product of an $S^4$ Euclidean instanton with any compact space.

The number 500 has become a sort of a code for the landscape, because this is the number of handles on a large Calabi-Yau manifold, but for now it is an arbitary guess.  It is still not certain whether the number of vacua in string theory is dense enough to account for the smallness of the cosmological constant, or even whether it is finite (it probably becomes finite with some bound on the size of the compact dimensions: compact systems in general have discrete spectra\footnote{See the talk by Douglas for further discussion of this and related issues.}).

The nuclear example has a hidden cheat, in that a small parameter has been put in by hand: the action for tunnelling of a nucleus through the Coulomb barrier is of order $Z_1 Z_2 (m_{\rm p}/m_{\rm e})^{1/2}$, and this stabilizes all the decays.  String theory has no such small parameter.  One of the key results of KKLT is that in some regions of moduli space there are a few small parameters that stabilize all decays (see also Ref.~\cite{Frey:2003dm}).  Incidentally, the stability of our vacuum is one reason to believe that we live near some boundary of moduli space, rather than right in the middle where it is particularly hard to calculate: most likely, states right in the middle of moduli space decay at a rate of order one in Planck units.

How trustworthy are the approximations in KKLT?  A skeptic could argue that there are no examples where they are fully under control.  Indeed, this is likely to inevitable in the construction of our vacuum in string theory.  Unlike supersymmetric vacua, ours has no continuous moduli that we can vary to make higher-order corrections parametrically small, and the underlying string theory has no free parameters.  It could be that our vacuum is one of an infinite discrete series, indexed by an integer which can be made arbitrarily large, and in this way the approximations made parametrically accurate, but in the KKLT construction this appears not to be the case: the flux integers and Euler number are bounded.  For future reference we therefore distinguish {\it series} and {\it sporadic} vacua, by analogy to finite groups and Lie algebras; perhaps other constructions, e.g. \cite{Saltman:2004jh},
give series of metastable nonsupersymmetric vacua.

The KKLT construction has something close to a control parameter, the supersymmetry breaking parameter $w_0$.  In an effective field theory description we are free to vary this continuously and then the approximations do become parametrically precise; in this sense one is quite close to a controlled approximation.  In specific models the value of $w_0$ is fixed by fluxes, and it is a hard problem (in a sense made precise in Ref.~\cite{Denef:2006ad}) to find vacua in which it is small.  Thus, for now the fourth prediction from the previous section, that string theory has enough vacua to solve the cosmological constant problem, is undecided and still might falsify the whole idea.

Underlying the above discussion is the fact that we still have no nonperturbative construction of string theory in any de Sitter vacua, as emphasized in particular in Refs.~\cite{Banks:2003es,Banks:2004xh}.  As an intermediate step one can study first supersymmetric AdS vacua, where we do understand the framework for a nonperturbative construction, via a dual \mbox{CFT}.  The KKLT vacua are built on such AdS vacua by exciting the system to a nonsupersymmetric state.  The KKLT AdS vacua are sporadic, but there are also series examples with all moduli fixed, the most notable being simply $AdS_5 \times S^5$, indexed by the five-form flux.  Thus far we have explicit duals for many of the series vacua, via quiver gauge theories, but we do not yet have the tools to describe the duals of the sporadic vacua~\cite{Silverstein:2003jp}.  The KKLT construction makes the prediction that there are $10^{O(100)}$ such sporadic CFTs --- a surprising number in comparison to the number of sporadic finite groups and Lie algebras, but indeed 2+1 dimensional CFTs appear to be much less constrained.  It may be possible to count these CFTs, even before an explicit construction, through some index; see Ref.~\cite{Douglas:2004zg} for a review of various aspects of the counting of vacua.

Beyond the above technical issues, there are questions of principle: are the tools that KKLT use, in particular the effective Lagrangian, valid?  In many instances these objections seem puzzling: the KKLT construction is little more than gluino condensation, where effective Lagrangian methods have long been used, combined with supersymmetry breaking, which can also be studied in a controlled way.  It is true that the KKLT construction, in combination with eternal inflation, is time-dependent.  However, over much of the landscape the scale of the time-dependence is well below the Planck scale, because the vacuum energy arises from a red-shifted throat, and so the landscape is populated in the regime where effective field theory is valid.

A more principled criticism of the use of effective Lagrangians appears in Refs.~\cite{Banks:2003es,Banks:2004xh}; I will try to paraphrase this here.  It is not precisely true that the nonsupersymmetric KKLT states (or any eternally inflating states) are excitations of AdS vacua.  That is, it is true locally, but the global boundary conditions are completely different.  Normally one's intuition is that the effective Lagrangian is a local object and does not depend on the boundary conditions imposed on the system, but arguments are given that this situation is different.
In particular one cannot tunnel among inflating states, flat spacetimes, and AdS states in any direction (for example, tunneling from eternal inflation to negative cosmological constant leads to a crunch); thus these are in a sense different theories.  This is also true from a holographic point of view: the dual Hamiltonians that describe inflating, flat, and AdS spaces will inevitably be completely different (as one can see by studying the high energy spectrum).  Is there then any reason to expect that constructions of an effective action, obtained from a flat spacetime S-matrix, have any relevance to an eternally inflating system?

I believe that there is.  The entire point of holography and AdS/CFT duality is that the bulk physics is emergent: we obtain the same bulk physics from many different Hamiltonians.  We can already see this in the AdS/CFT context, where many different quiver gauge theories, even in different dimensions, give the same IIB bulk string theory, and local experiments in a large AdS spacetime are expected to give the same results as the same experiments in flat spacetime.  Thus there is no argument in principle that these do not extend to the inflating case.  Also, while holography does imply some breakdown of local field theory, it does so in a rather subtle way, as in phase correlations in Hawking radiation.  By contrast, the expectation value of the energy-momentum tensor in the neighborhood of a black hole (\mbox{i.e.} the total flux of Hawking radiation) appears to be robust, and the quantities that enter into to construction of string vacua are similar to this.

However, for completeness we mention the possible alternate point of view~\cite{Banks:2005bm}: that the 
landscape of metastable dS vacua has no nonperturbative completion, or it does have one but is experimentally ruled out by considerations such as those we will discuss.  Instead there is a completely separate sector, consisting of theories with finite numbers of states, and if these lead to emergent gravity it must be in a stable dS spacetime.

\subsection{Phenomenological issues}

Thus far we have dwelt on the cosmological constant, but the string landscape implies that other constants of nature will be environmental to greater or lesser extents as well.  In this section we discuss a few such parameters, especially those which appear to be problematic for one reason or another. 

\subsubsection{$\theta_{\rm QCD}$}  

Why is $\theta_{\rm QCD}$ of order $10^{-9}$ or less?  This strong CP problem has been around for a long time in gauge theory, and several explanations have been proposed --- an axion, a massless up quark, and models based on spontaneous CP violation.  However, it has been argued that none of these are common in the string landscape; for example, the first two require continuous symmetries with very tiny explicit breakings, and this appears to require fine tuning.  Further, it is very hard to see any anthropic argument for small $\theta_{\rm QCD}$; a larger value would make very little difference in most of physics.  Thus we would conclude that the multiverse is full of bubbles containing observers who see gauge theories with large CP-violating angles, and ours is a one-in-a-billion coincidence~\cite{Banks:2003es}.

Of course, this is a problem that is to some extent independent of string theory: the axion, for example, has always been fishy, in that one needed a global symmetry that is exact {\it except for} QCD instantons.  The string landscape is just making sharper an issue that was always there. 

String theory does come with a large number of potential axions.  In order that one of these solve the strong CP problem it is necessary that the potential energy from QCD instantons be the dominant contribution to the axion potential; any non-QCD contribution to the axion mass must be of order $10^{-18}\,{\rm eV} \times (10^{16}\, {\rm GeV}/f_{\rm a})$ or less; see e.g. \cite{Conlon:2006tq} for a recent discussion.  This is far below the expected scale of the moduli masses, so appears to imply a substantial fine tuning (even greater than the direct tuning of $\theta_{\rm QCD}$) and so rarity in the landscape.

However, the landscape picture also suggests a particular solution to this problem.  In order to obtain a dense enough set of vacua, the compact dimensions must be topologically complex, again with something around 500 cycles.  Each cycle gives rise to a potential axion, whose mass comes from instantons wrapping the cycle (we must exclude would-be axions which also get mass from other sources, such as their classical coupling to fluxes).  Generically one would expect some of these cycles to be somewhat large in string units; for example, one might expect the whole compact space to have a volume that grows as some power of the number of handles.  The axions, whose masses go as minus the exponential of the volume, would be correspondingly light.  Thus, compactifications of large topological complexity may be the one setting in which the QCD axion {\it is} natural, the smallness of $\theta_{\rm QCD}$ being an indirect side effect of the need for a small cosmological constant.  More generally, it will be interesting to look for characteristic properties of such topologically complex compactifications.

This example shows that even with anthropic selection playing a role, mechanism will surely also be important.

\subsubsection{The baryon lifetime}  

This is a similar story to $\theta_{\rm QCD}$~\cite{Banks:2003es}: as far as we understand at present, the baryon lifetime is longer than either anthropic argument or mechanism can account for, so that bubbles with such long-lived baryons would be rare in the multiverse.  This problem is lessened if supersymmetry is broken at high energy.  This is an significant challenge to the landscape picture: it is good to have such challenges, eventually to sharpen, or to falsify, our current understanding.

\subsubsection{The dark energy parameter $w$}

A naive interpretation of the anthropic principle would treat the dark energy equation of state parameter $w$ as arbitrary, and look for anthropic constraints.  However, in the string landscape a simple cosmological constant, $w = -1$, is certainly favored.  With supersymmetry broken, the scalar potential generically has isolated minima, with all scalars massive.  In order to obtain a nontrivial equation of state for the dark energy we would need a scalar with a mass of order the current Hubble scale.  Our discussion of axions indicates a mechanism for producing such small masses, but it would be rather contrived, for no evident reason, that the mass would be of just the right scale as to produce a nontrivial variation in the current epoch.

\subsubsection{Three generations}

Three generation models appear to be difficult to find in string theory.  A recent paper quantifies this~\cite{Gmeiner:2005vz}: in one construction they are one in a billion, even after taking into account the anthropic constraint that there be an asymptotically free group so that the long distance physics is nontrivial.  It is then a puzzle to understand how we happen to live in such a vacuum.  One conjecture is that all constructions thus far are too special, and in the full landscape three generations is not rare.  Again, explaining three generations is equally a problem for any hypothetical alternate selection mechanism --- another challenge to sharpen our understanding.

\subsubsection{$Q$}

I am not going to try to discuss this parameter in detail; I am only going to use it to make one rhetorical point.  The anthropic bound on $Q$, which is the normalization of the primordial temperature fluctuations, has been quoted as~\cite{Tegmark:1997in} 
\begin{equation}
10^{-6} < Q < 10^{-4}\ ,
\end{equation}
and it is interesting that the observed value is in the middle, not at either end.  What would we expect from the landscape? 

A string theorist would note that the anthropic bound is on $\rho_V Q^{-3}$~\cite{Weinberg:1987dv}, and so by making $Q$ a factor of 10 larger we can multiply $\rho_V$ by 1000, and there will be many more vacua with this larger value of $Q$.  A cosmologist would note that a smaller $Q$ would imply a flatter potential and so more inflation, and therefore much more volume and many more galaxies.  Thus the cosmologist and the string theorist agree that we should be on the end of the anthropic range, but they disagree on which end.

This is a caricature, of course --- there are other considerations, and model-dependencies~\cite{Banks:2003es,Graesser:2004ng,Feldstein:2005bm,Garriga:2005ee,Tegmark:2005dy}.  I use it to make two points: first, it is a puzzle that we are in the middle of the anthropic range, yet another thing to understand.  Second, the string theorist and the cosmologist each look at part of the measure, but it is clear that we are far short of the whole picture.
(For reviews of the counting and the volume factors see Refs.~\cite{Douglas:2004zg} and~\cite{Vilenkin:2006qf} respectively.)

\subsubsection{$\rho_V$}

Can we understand understand the number 283, as in 
\begin{equation}
\rho_V = e^{-283.2} M_{\rm P}^4\ ?
\end{equation}
I quote it in this way, as a natural log, to emphasize that we are to think about it completely free of all priors (such as the fact that we have ten fingers).   Thus, there may be an anthropic relation between $\rho_V / M_{\rm P}^{4}$ and $M_{\rm weak} / M_{\rm P}$, for example, but we should not make any assumption about the latter.
It should be possible to calculate the number 283, at least to some accuracy.  We know that it has to be big, to get enough bits and cycles, but why is 100 not big enough, and why is 1000 not better?

One possibility, the best from the point of view of string theory, is that $\rho_V / M_{\rm P}^{4}$ has its original purely in microphysics; that it, that it is close to the smallest attainable value, set by the density of the discretuum.  The other extreme is that it is almost purely anthropic --- that 283, plus or minus some uncertainty, really gives the best of all attainable worlds, and any attempt to vary parameters to give a larger or smaller value makes things worse.  Certainly, knowing where we sit between these two extremes is something that we must eventually understand in a convincing way.

\subsubsection{Other questions}

An obvious question is whether we can understand the supersymmetry-breaking scale (see~\cite{Dine:2005yq} and references therein).  Is low energy supersymmetry, or some alternative~\cite{Arkani-Hamed:2004fb,Fox:2005yp}, favored?  Will we figure this out before the LHC tells us?

Another potentially telling question~\cite{Aguirre:2004qb}: are there more coincidences like the cosmic coincidence of $\rho_V$, such as the existence of two different kinds of dark matter with significant densities?

\subsection{What is string theory?}

Of course, this is still the big question.  We have learned in recent years that the nonperturbative construction of a holographic theory is very sensitive to the global structure of spacetime.  Thus, the current point of view, the chaotically inflating multiverse, casts this question in a new light.  It is also another example of how the landscape represents productive science: if we ignore this lesson, ignore chaotic inflation, we may be trying to answer the wrong question.

Before addressing the title question directly, let us discuss one way in which it bears upon the previous discussion.  We touched briefly on the issue of the measure.  This has always been a difficult question in inflationary cosmology.  Intuitively one would think that the volume must be included in the weighting, since this will be one factor determining the total number of galaxies of a given type.  However, this leads to gauge dependence~\cite{Linde:1993xx}
and the youngness paradox~\cite{Guth:2000ka}.  Further, this would imply that the vacuum of highest density plays a dominant role, whereas the de Sitter entropy would suggest almost the opposite, that when the system is in a state of high vacuum energy it has simply wandered into a subsector of relatively few states.  Further, the idea of counting separately regions that are out of causal contact is contrary to the spirit of the holographic principle.

There have been attempts to modify the volume weighting to deal with some of the paradoxes (for a recent review see Ref.~\cite{Vilenkin:2006qf}), but as far as I know none as yet take full advantage of the holographic point of view, and none is widely regarded as convincing.  Providing a compelling understanding of the measure is certainly a goal for string theory.  It is possible that this can be done by some form of holographic reasoning, even without a complete nonperturbative construction.  It is perhaps useful to recall Susskind's suggestion, that the many worlds of chaotic inflation are the same as the many worlds of quantum mechanics.  This can be read in two directions: first, that chaotic inflation is the origin of quantum mechanics --- this seems very ambitious; second, that the many causal volumes in the chaotic universe should just be seen as different states within the wavefunction of a single patch --- this is very much in keeping with holography.  It is also interesting to note that in the stochastic picture presented in Ref.~\cite{Linde:1993xx} several probability measures are considered; some have a youngness paradox, but at least one seems to connect with the Hartle-Hawking and tunneling wavefunctions, and possibly with a thermodynamic picture.

Now, what is the nonperturbative construction of these eternally inflating states?  The lesson from AdS/CFT is that the dual variables that give this construction live at the boundary of spacetime.  In the context of eternal inflation, the only natural boundaries lie to the future, in open FRW universes (and possibly also in time-reversed universes to the past)~\cite{Susskind:2003kw,Freivogel:2004rd,Bousso:2005yd}.

This is much like AdS/CFT with timelike infinity replacing spatial infinity, and so it suggests that time will be emergent.  Let us interpose here one remark about emergent time (see also the presentations by Seiberg and by Maldacena at this meeting).  Of course in canonical general relativity there is no time variable at the start, it emerges in the form of correlations once the Hamiltonian constraint is imposed.  This sounds like emergent time, but on the other hand it is just a rewriting of the covariant theory, and one would expect emergent time to be something deeper.  

To see the distinction between emergent time in these two senses let us first review emergent gauge symmetry.  In some condensed matter systems in
which the starting point has only electrons with short-ranged interactions,
there are phases where the electron separates into a new fermion
and boson~\cite{D'Adda:1978uc,Baskaran:1987my},
\begin{equation}
e(x) = b(x) f^\dagger(x)\ .
\end{equation}
However, the new fields are redundant: there is a gauge transformation
\begin{equation}
b(x,t) \to e^{i \lambda(x,t)} b(x,t)\ , \quad f(x,t) \to  e^{i \lambda(x,t)}  f(x,t)\ , 
\end{equation}
which leaves the physical electron field invariant. This new gauge invariance is clearly emergent: it is completely invisible in terms of the electron field appearing in the original
description of the theory (this ÔstatisticalÕ gauge invariance is not to be confused with the ordinary electromagnetic gauge invariance, which does act on the electron.)  Similarly, the gauge theory variables of AdS/CFT are trivially invariant under the bulk diffeomorphisms, which
are entirely invisible in the gauge theory (the gauge theory fields do
transform under the asymptotic symmetries of $AdS_5 \times S^5$, but these are
ADM symmetries, not gauge redundancies).  

Thus, in the case of emergent time we look for a description of the theory in which time reparameterization invariance is invisible, in which the initial variables are trivially invariant.  It is not a matter of solving the Hamiltonian constraint but of finding a description in which the Hamiltonian constraint is empty.  Of course we can always
in general relativity introduce a set of gauge-invariant observables by
setting up effectively a system of rods and clocks, so to this extent the
notion of emergence is imprecise, but it carries the connotation that the
dynamics can be expressed in a simple way in terms of the invariant
variables.  The AdS/CFT duality solves this problem by locating the variables at spatial infinity, and in the present context the natural solution would be to locate them at future infinity.  That is, there some dual system within which one calculates directly the outgoing state in the FRW patches, some version of the Hartle-Hawking wavefunction perhaps.  To access our physics in a nonsupersymmetric and accelerating bubble would then require some holographic reconstruction as in the bulk of AdS/CFT.  Certainly such a picture would cast a very different light on many of the questions that we have discussed; it does suggest a possible mechanism for `post-selection' of the cosmological constant.

It would be useful to have a toy model of emergent time.  The problem with the string landscape is that all states mix, and one has to deal with the full problem; is there any isolated sector to explore?

\setcounter{equation}{0}
\section{Conclusions}

A few closing thoughts:
\begin{itemize}
\item
The extent to which first principles uniquely determine what we see in nature is itself a question that science has to answer.  Einstein asked how much choice God had, he did not presume to know the answer.
\item
That the universe is vastly larger than what we see, with different laws of physics in different patches, is without doubt a logical possibility.  One might argue that even if this is true it is forever outside the domain of science, but I do not think it is up to us to put a priori bounds on this domain.  Indeed, we now have five separate lines of argument (the predictions near the end of Sec.~1) that point in this direction.  Our current understanding is not frozen in time, and I expect that if this idea is true (or if it is not) we will one day know.
\item
A claim that science is less predictive should be subjected to a correspondingly higher level of theoretical skepticism.  Our current picture should certainly be treated as tentative, at the very least until we have a nonperturbative formulation of string theory.
\item
The landscape opens up a difficult but rich spectrum of new questions, \mbox{e.g.}~\cite{Intriligator:2006dd}.
\item
There are undoubtedly many surprises in the future. 
\end{itemize}
Let me close with a quotation from Dirac:
\begin{quote}
One must be prepared to follow up the consequences of theory, and feel that one just has to accept the consequences no matter where they lead.  
\end{quote}
and a paraphrase:
\begin{quote}
One should take seriously all solutions of one's equations.
\end{quote}
Of course, his issue was a factor of two, and ours is a factor of $10^{500}$.

\vspace{5mm}

\noindent {\bf Acknowledgments}:
I would like to thank Nima Arkani-Hamed, Lars Bildsten, Raphael Bousso, Clifford Burgess, Michael Dine, Michael Douglas, Fernando Quevedo, Eliezer Rabinovici, and Eva Silverstein for helpful discussions.  This work was supported in part by NSF grants PHY99-07949 and PHY04-56556.

\end{document}